# SISTEM PENGAMBILAN KEPUTUSAN PENANGANAN BENCANA ALAM GEMPA BUMI DI INDONESIA

Spits Warnars Harco Leslie Hendric (s.warnars@mmu.ac.uk)
Department of Computing and Mathematics, Manchester Metropolitan University
John Dalton Building, Manchester M15 6BH, United Kingdom, Tel +44(0)1612472000, Fax +44(0)1612476390

**Abstract:**
After Aceh's quake many earthquakes have struck Indonesia alternately and even other disasters have been a threat for every citizen in this country. Actually an everyday occurrence on earth and more than 3 million earthquakes occur every year, about 8,000 a day, or one every 11 seconds in Indonesia there are 5 to 30 quakes prediction everyday. Government's responsibility to protect the citizen has been done by making National body of disaster management. Preparing, saving and distribution logistic become National body of disaster management's responsibility to build information management. Many law's products have been produced as a government's responsibility to give secure life for the citizen. We can not prevent them totally, we have to learn to live with them and need to be prepared all the time, need to learn how to mitigate risk of losses in such events by managing crisis and emergencies correctly. After disaster happens respond must be rapidly and at an optimal level to save lives and help to victims. DSS is information technology environment which can be used to help human in order to learn from past earthquake, record it, learn and plan for future mitigation and hope will reduce the disaster risk in the future. Using web technology for DSS will give value added where not only make a strategic decision for the decision maker, but for others who need national earthquake information like citizen, scholars, researches and people around the world.

**Kata Kunci: Sistem Pengambilan Keputusan, Penanganan Bencana Alam Gempa Bumi, E-Government, Data Warehouse**

## 1. PENDAHULUAN

Jakarta sebagai pusat pemerintahan Indonesia pada tanggal 2 september dikagetkan dengan gempa bumi yang berkekuatan 7.3 pada skala righter yang terjadi yang merusak Jawa Barat dan sekitarnya dengan isu adanya ancaman Tsunami namun tidak terbukti, telah memakan korban jiwa sebanyak hamper 80 orang. Bandingkan dengan gempa yang terjadi baru-baru ini di provinsi Sichuan, China yang berkekuatan 7.9 pada skala righter pada 12 Mei 2008 yang memakan korban 40000 orang lebih. Jika anda mampir ke www.bmg.go.id anda akan mendapatkan data-data gempa per hari yang terjadi di seluruh Indonesia. Belum juga hilang dari ingatan kita dengan kejadian-kejadian bencana alam gempa bumi sebelumnya yang menghantam Indonesia:
1. Di Aceh dan sekitarnya yang berkekuatan 9.1 pada skala righter yang menyebabkan gelombang Tsunami yang terjadi pada 26 Desember 2004 memakan korban jiwa lebih dari 170 ribu jiwa.
2. Di Pulau Nias dan sekitarnya yang terjadi pada 28 Maret 2005 yang berkekuatan 8.7 pada skala righter memakan korban jiwa lebih dari 1000 orang.

3.  Di Yogyakarta pada hari Sabtu, 27 Mei 2006 yang berkekuatan 5.9 pada skala righter dimana gempa mulai terjadi selama kurang lebih 1 menit pada pukul 05.53 pagi telah memakan korban jiwa lebih dari 5000 jiwa.

Banyak orang yang memberikan pendapatnya berkenaan dengan bencana alam gempa bumi ini. Menurut pendapat beberapa orang bencana alam ini merupakan karya ajaib sang Maha kuasa dan tanda-tanda alam yang tidak dapat kita tolak. Bagi beberapa orang ini merupakan peringatan dari yang Maha Kuasa agar manusia di muka bumi ini selalu mengagungkan Namanya dan menjauhi perbuatan jahat. Ada juga yang berpendapat bahwa ini semua adalah hukuman bagi orang jahat.

Semua pendapat diatas tentang bencana alam gempa bumi yang selalu dikaitkan dengan ilmu agama dan KeTuhanan tersebut sah-sah saja. Namun terlepas dari semua pendapat diatas tersebut bencana alam gempa bumi ini janganlah membuat kita sebagai makhluk yang diciptakan cerdas dan serupa dengan gambaranNya hanya bisa diam saja dan menerima apa yang merupakan kehendakNya. Walaupun tepat dari sudut pandang keTuhanan kita harus hidup lebih baik lagi dan merupakan sebagai peringatan bagi yang masih hidup, namun tidak tertutup kemungkinan yang Maha Kuasa telah memberikan kepada kita otak yang harus dapat digunakan dan lagi cerdas adanya. Baik secara langsung atau tidak langsung dan pasti yang Maha Kuasa ingin agar kita menggunakan tingkat kecerdasan otak kita untuk membantu sesama kita. Bencana alam gempa bumi ini merupakan hak dari sang Maha Kuasa dan kewajiban kita adalah bagaimana memanfaat kan kecerdasan yang telah diberikanNya untuk dapat menyelamatkan hak hidup manusia terlepas dari takdir yang mengatakan nyawa manusia berada dalam tangan yang Maha Kuasa.

Saat ini yang namanya bencana alam gempa bumi ini belum ada ilmu pengetahuan yang dapat memprediksi kapan gempa akan terjadi dan dimana akan terjadi. Kalapun prediksi yang ada saat ini berdasarkan bencana alam yang merupakan bencana alam yang terjadi secara periodik dan dapat dibaca dari pergerakan arah angin. Saat ini yang dapat diprediksi adalah bagaimana penanganan akibat dari yang ditimbulkan oleh bencana alam, seperti akan timbulnya bencana tsunami jika gempa terjadi di lautan. Ilmu untuk memprediksi ini dapat digunakan untuk membantu penanganan paska gempa dalam rangka mengurangi kerugian,baik kerugian nyawa maupun kerugian harta benda.

Dalam dunia komputer ilmu untuk memprediksi dapat dibuatkan sebuah sistem informasi yang disebut sebagai *Decision Support System* atau jika diterjemahkan dalam bahasa Indonesia adalah sistem penunjang keputusan. Penting bagi bangsa Indonesia untuk dapat menggunakan kecerdasan yang telah diberikan oleh yang Maha Kuasa untuk membantu sesama dan mengurangi kerugian yang diakibatkan oleh bencana alam gempa bumi.

Tulisan ini akan mencoba membahas bentuk DSS yang paling cocok dan sesuai dengan situasi dan kondisi Indonesia. Membicarakan aspek-aspek pendukung terwujudnya DSS dan mencoba melihat dari riset-riset atau benchmark yang telah ada dan berkorelasi dengan tulisan ini. Diharapkan dengan ide dibuatnya DSS ini bangsa Indonesia akan mempunyai sebuah sistem yang akan membantu bangsa ini untuk tetap melanjutkan pembangunan negara dan terus bangkit dari keterpurukan. DSS ini diharapkan dapat dipakai oleh semua tingkatan level masyarakat,mulai dari masyarakat kecil sampai pejabat menteri hingga presiden. Dengan demikian kerugian yang diakibatkan oleh gempa bumi dapat diminimalisasi.

Yang salah dari bangsa ini adalah kita tidak pernah belajar dari kesalahan-kesalahan dan dari knowledge yang telah ada. Knowledge penanganan baik korban jiwa maupun kerugian material gempa bumi mulai dari Aceh, Nias dan Yogya bahkan mulai dari bencana alam terdahulu tidak pernah terdata secara terpusat. Sehingga kita selalu melakukan kesalahan-kesalahan yang sama, korban jiwa dan kerugian material yang selalu sama bahkan meningkat.

Ide dari korban jiwa yang banyak dari korban gempa Yogya adalah kurangnya tenaga medis yang turun ke lapangan saat setelah terjadi gempa. Banyak korban yang tidak sempat tertolong karena kurangnya tenaga medis dan ini merupakan point paling penting yang harus dicermati dengan dibuatnya DSS ini. Diharapkan dengan dibuatnya DSS ini pengambil keputusan dalam hitungan menit dapat menentukan luas area terkena bencana gempa bumi, berapa kemungkinan korban yang ada, berapa tenaga medis yang ada di sekitar bencana gempa bumi tersebut yang dapat dilibatkan ? Kalau kurang maka dalam hitungan menit tersebut kita dapat menyampaikan SOS kepada dunia internasional.

Sampai saat ini para pengambil keputusan di level pemerintah tidak didukung dengan informasi yang terpusat dan dapat dipercayai oleh semua orang. Kalaupun ada didapatkan dari berbagai sumber yang tidak bisa dipastikan dari mana asalnya dan akhirnya keputusan yang diambil hanya asal tebak. Sehingga tidak ada lagi seorang menteri ketika ditanya kapan makanan untuk para pengungsi akan sampai dan dijawab setidaknya 3 sampai 4 hari. Jawaban seperti ini adalah jawaban yang tidak manusiawi, bagaimana jawaban 3 sampai 4 hari itu didapatkan itupun kalau sesuai, yang terjadi di lapangan kadang melebihi waktu tersebut. Tidak masuk akal juga jika bantuan dilakukan dalam tempo waktu yang sangat lama tersebut, jarak Yogya dari pusat pemerintahan negara Indonesia jika kita naik bis malam hanyalah 7 jam dan apabila naik pesawat terbang hanya 1 jam. Bagaimana waktu 1 atau 7 jam tersebut dibandingkan dengan 3 sampai 4 hari sangatlah tidak masuk akal. Seharusnya pemerintah mempunyai sebuah sistem DSS penanganan gempa yang dapat mengatur distribusi dalam waktu sesuai dengan waktu jarak transportasi standar dan tidak dilebih-lebihkan.

Sudah saatnya pola pikir para pengambil keputusan di negara kita ini berpihak kepada rakyatnya dan sayang kepada rakyatnya dan tidak selalu menyiksa rakyatnya. Yang hanya bisa dilakukan oleh para pengambil keputusan di negara kita ini hanyalah menaikkan harga yaitu :

1. Menaikkan harga minyak bumi, gas.
2. Menaikkan harga bayar alias gaji para pengambil keputusan.
3. Menaikkan harga pinjaman alias berhutang.

Bagaimana negara ini mau maju dengan pola pikir para pengambil keputusan yang jelas-jelas berkelakuan monopoli. Bandingkan dengan harga pulsa handphone yang dari waktu ke waktu semakin turun harganya. Dengan banyaknya provider nomor handphone ini maka satu provider akan bersaing dengan provider lainnya untuk dapat merebut hati konsumen. Bahkan mungkin ke depan konsumen tidak perlu membayar jika menggunakan pulsa handphone ini. Kembali ke soal monopoli, jelas tidak adanya iklim persaingan menyebabkan yang hanya bisa dilakukan hanyalah menaikkan harga dan karena monopoli mau tidak mau rakyat dengan sangat sangat terpaksa dan bercucuran air mata menerima keputusan tersebut. Rakyat tidak ada pilihan lain beda dengan iklim persaingan, orang dapat memilih sesuai dengan selera apakah dimulai dari harga murah baru bicara kualitas ataukah dimulai dari kualitas baru bicara harga.

Kebijakan menaikkan hutang jelas-jelas tidak masuk akal dan terlalu berlebihan. Jelas-jelas rakyat diajarkan untuk hidup berhutang dan berhutang itu biasa-biasa saja. Tuduhan beberapa politisi bahwa pemerintahan SBY-JK hanya memberantas korupsi hanyalah sebatas "show performance" dan terkesan tidak menyeluruh menjadi benar. Masih banyak uang rakyat yang dipungut "paksa" dari rakyat setiap bulannya tidak jelas kemana ! Makanya banyak orang enggan untuk membayar pajak dengan benar karena tahu uang pajaknya tidak dimanfaatkan dengan benar. Sudah jadi hal umum di masyarakat kita bahwa korupsi adalah hal yang biasa dan merupakan kekhilafan manusia. Sudah jadi hal umum di masyarakat kita kalau mau beres harus ada uang pelicin, kalau tidak maka akan dipersulit, beda dengan maling ayam atau sendal yang harus meninggal dihakimi oleh massa karena mencari sesuap nasi.

Tidak mungkin sebuah keluarga dapat hidup tenang kalau kehidupan ekonominya ditunjang oleh hutang, makan tidak tenang, tidur tidak tenang memikirkan bunga hutang yang makin meningkat. Tapi hal ini berbeda dengan apa yang dialami oleh para pengambil keputusan negara ini, ciri-ciri rumah tangga yang berhutang tidak terlihat pada para pengambil keputusan negara ini. Manusia yang lahir di negara ini lahir menjadi seorang budak yang mempunyai hutang, karena mempunyai hutang jelas manusia yang lahir di negara ini adalah budak yang tidak dapat secara bebas menentukan hidupnya karena dibatasi oleh hutang-hutangnya. Sampai berapa lama lagi generasi berikutnya akan membayar hutang-hutang negara ini. Jangan salahkan sebagian rakyat negeri ini yang melirik ke tempat lain yang tidak melahirkan anak-anaknya sebagai manusia yang berhutang. Jangan Salahkan sebagian rakyat Papua yang lari ke Australia, karena mereka sebagai manusia mempunyai hak hidup dasar untuk tidak menjadi manusia yang mempunyai hutang.

## 2. IDENTIFIKASI PERMASALAHAN

Pemerintah mempunyai tanggung jawab untuk menyatakan keadaan darurat dimulai dari sebelum bencana, pada saat bencan dan setelah bencana [Pelaksana 2007, Presiden 2007]. Setiap kejadian bencana alam gempa bumi menciptakan ancaman baik untuk kehidupan dan harta benda dan menjadi tanggung jawab pemerintah untuk melakukan penanggulangan sebagai hak setiap warga negara untuk mendapatkan jaminan perlindungan. Belajar dari bencana gempa bumi yang lampau, merekamnya, pelajari dan rencanakan untuk prediksi penanggulangan dan berharap akan mengurangi resiko ancaman di masa yang akan datang.

Pemerintah sebagai pengambil keputusan yang menentukan tingkat dan status bencana [Presiden 2007, Presiden 2008b] tidak pernah didukung dengan bantuan teknologi informasi yang dapat membantu dalam membuat keputusan yang baik dan cepat dengan informasi yang benar dan dapat dipercaya. Penundaan pembuatan keputusan akan menambah jumlah korban yang terlantar terutama orang tua, perempuan dan anak-anak. Jika pemerintah mendapatkan informasi sebagai masukan untuk keputusan, apakah itu informasi yang benar dan dapat dipercaya atau seakan-akan hanya informasi yang hanya menyenangkan pemerintah ? Bagaimana dengan data-data bencana alam gempa bumi yang sudah pernah terjadi, apakah data-data tersebut diabaikan saja dan tidak menjadi pelajaran untuk kedepan ? Bagaimana dengan kegagalan dan keberhasilan bencana alam yang lampau ? Apakah kita sudah belajar dari sejarah yang kadang-kadang menjadi hal yang menarik dan informative dan membuat kita semakin tahu mana yang kurang berhasil dan apa yang sudah berhasil ?

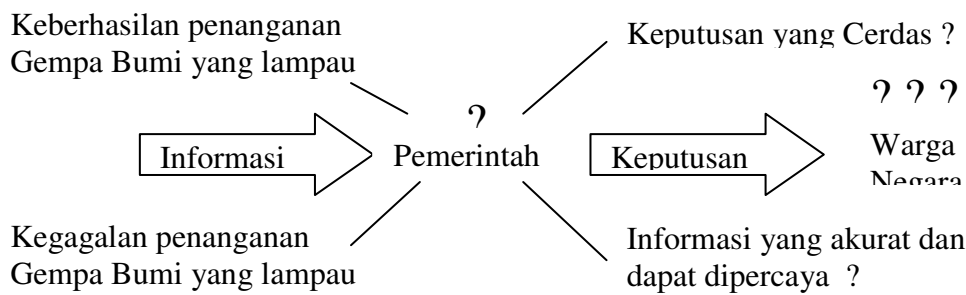

Gambar 1. Pemerintah ketika membuat keputusan penanggulangan bencan alam gempa bumi

Apakah setiap warga Negara mengetahui data-data tentang bencana alam gempa bumi dari satu sumber yang sama dan dapat dipercaya atau hanya berdasarkan isu-isu yang ada di majalan dan surat kabar ? Pemerintah kita mempunyai sebuah website yang mengarah kesana yaitu www.bakornaspb.go.id, namun jadi pertanyaan apakah situs ini masih dipelihara atau tidak ? Sekalipun situs ini didukung dengan fasilitas forum sebagai komunikasi untuk warga, namun jika melihat isi forum di situs, maka hany akan didapatkan sebuah pesan dari seorang warga Negara yang putus asa dengan informasi yang selalu dan tidak akan pernah didapat ! Jika begitu ada apa, apakah badan penanggulangan bencana menjalankan tugas mereka untuk memberikan tanggungjawab yang dapat dipertanggungjawabkan sebagaimana mereka sebagai pegawai negeri yang mendapatkan gaji dari pajak yang dipungut dari warga ? [Presiden 2008b] Dimana tanggung jawab untuk memberikan perlindungan kepada warga, mendidik dan melatihnya dan menanggulangi bencana alam gempa bumi ?[Presiden 2007]

Teknologi informasi dapat dan harus digunakan oleh pemerintah untuk menjamin hak-hak warga negara untuk dilayani oleh pemerintah dan tidak hanya menghukum ketika tidak atau telat membayar pajak. EGovernment sebagai hal yang telah lama didengungkan tidak jelas kemana arahnya, walaupun ada juga sebagian departemen dan pemerintah daerah menggunakannya sebagai alat komunikasi untuk warga yang dilayaninya. Pelaksanaan EGovernment harus transparan dan dapat dikembangkan menjadi *Government to citizen* (G2C), *Government to business* (G2B) dan *government to government* (G2G)[Turban 2005]. Harus ada keseriusan dari pemerintah untuk membentuk sebuah sistem yang dapat membantu mereka sebagai pengambil keputusan, warganya, generasi yang akan datang dengan transparansi dan dapat dipertanggungjawabkan. Jangan biarkan ilmu dan pengetahuan tentang bencana alam gempa bumi hanya tersimpan di dalam isi kepala aparat pemerintah dan tidak diinformasikan dan dilaksanakan di masyarakat sebagai pertanggung jawaban aparat pemerintah sebagai pelayan masyarakat. Warga harus dapat mengakses informasi setiap saat mereka butuhkan, dimana saja dan bahkan dunia internasional dapat mengaksesnya secara detail dan lengkap.

Sistem pengambilan keputusan adalah pendekatan yang dapat dipakai sebagai dukungan teknologi informasi yang dapat membuat keputusan yang cepat dan akurat [Coskun 2006]. Sistem pengambilan keputusan dapat menyimpan data lampau,mengelolanya dan menggunakannya sebagai bagian untuk membuat keputusan dan dapat digunakan untuk sistem pengambilan keputusan yang cerdas. Sistem pengambilan keputusan tidak hanya untuk pengambil keputusan, namun dapat digunakan untuk semua warga negara, ilmuwan, peneliti dan seluruh dunia dapat mengaksesnya dengan data yang benar dan akurat, sesuai dengan batasan-batasan data yang dapat diakses.

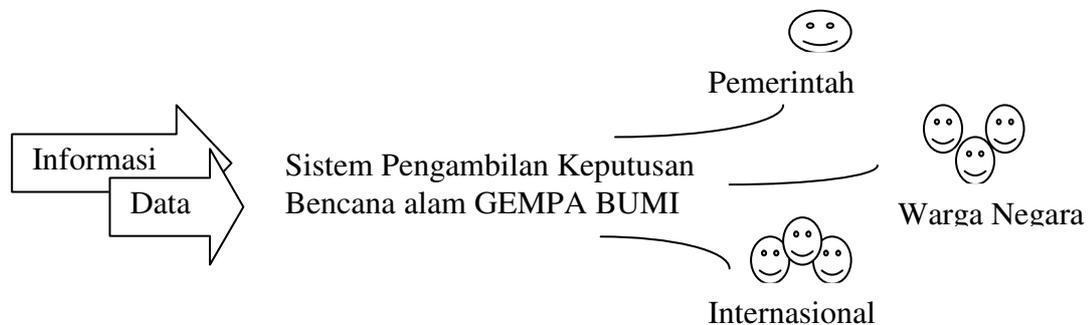

Gambar 2. Sistem Pengambilan keputusan untuk semua personal

Sistem pengambilan keputusan sebagai EGovernment implementasi dapat memuaskan warga sebagai pembayar pajak sebagai perwujudan tanggung jawab pemerintah untuk melindungi keselamatannya, hidup dan bisnis yang dijalankan. Setiap warga dapat mengetahui standar keselamatan yang ada dan akan atau yang telah dinyatakan oleh pemerintah dalam membantu warga untuk melindungi nyawa, harta benda dan bisnis yang dijalankan. Dengan jelasnya standar yang ada maka warga dapat menjalankan perekonomian dengan tenang dan keadaan bencana alam gempa bumi diharapkan tidak akan mengganggu perekonomian yang ada. Dalam keadaan bencana alam gempa bumi, perekonomian lokal yang terhimbas bencana tersebut dapat cepat teratasi dalam waktu singkat dan bahkan bisa diharapkan tidak mempengaruhi perekonomian yang ada. Didalam penanganan perekonomian penanganan isu-isu buruk diharapkan tidak mempengaruhi dikarenakan adanya isu-isu positip yang dapat dipercaya, satu sumber dan dikelola dengan benar oleh pemerintah dan diterima masyarakat luas bahkan dunia internasional. Sehingga perekonomian Indonesia tetap stabil dan tidak terpengaruh oleh isu-isu negatip berkenaan dengan bencana alam gempa bumi.

## 3. ARSITEKTUR SISTEM PENGAMBILAN KEPUTUSAN

Sistem pengambilan keputusan untuk penanggulangan bencana alam gempa bumi ini adalah kombinasi software, hardware dan teknologi yang menyediakan informasi sebagai dukungan untuk membuat keputusan. Sistem pengambilan keputusan penanggulangan gempa bumi dapat diimplementasikan sebagai EGovernment aplikasi yang dibuat dengan web teknologi. Dengan web teknologi untuk pembuatan EGovernment maka sistem dapat diakses online, dari mana saja, kapan saja dengan transparansi dan dapat dipercaya. Sistem pengambilan keputusan penanggulangan gempa bumi dapat diimplementasikan dengan
1) Pemrograman client seperti HTML, Javascript or Jscript, Code Style Sheet, and Java Applet.
2) Pemrograman Server seperti ASP, PHP, JSP, coldfusion or CGI
3) Database dapat menggunakan MySQL, Oracle, SqlServer, atau postgrace.
Pemilihan pemrograman server dan teknologi database tergantung pada kesepakatan, pilihan antara free open software atau license software. Untuk pelayanan web hosting service banyak ISP (Internet Service Provide) yang dapat dipilih dengan berbagai macam harga dan fasilitas yang disediakan dan pasti tergantung teknologi software yang digunakan dan budget yang disiapkan.
Multimedia informasi akan sangat bermanfaat untuk meningkatkan tampilan sistem, semua multimedia informasi seperti teks, bunyi, gambar, video dapat disertakan. Aplikasi dan database akan dibagi menjadi 2 server untuk meningkatkan performance dan untuk keamanan akan dibutuhkan firewall untuk mengawasi setiap akses. Khususnya untuk server database akan mempunyai 2 macam database yaitu database transaksi dan data warehouse sebagai struktur database yang tidak normal untuk analisa multi dimensi, menyimpan informasi, data histori, data eksternal sebagai hypercubes [Ponniah 2001].

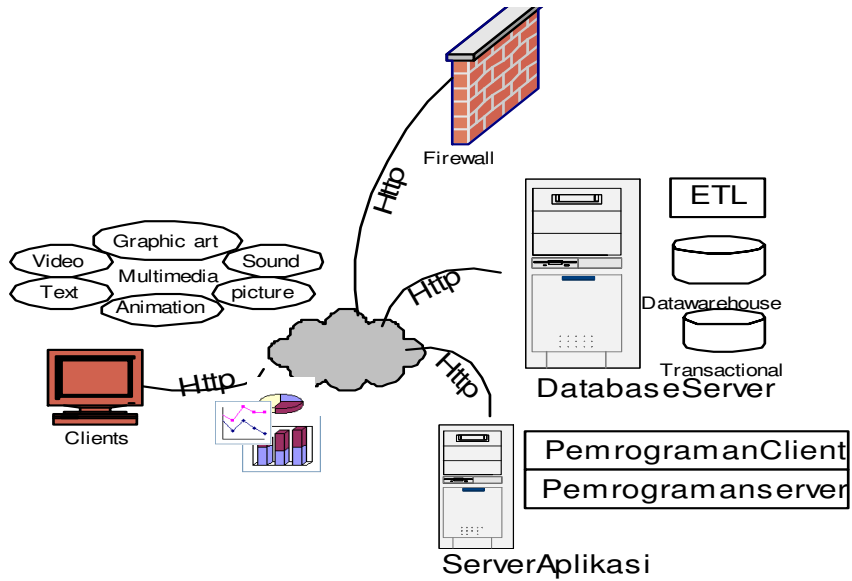

gambar 3. Arsitektur sistem pengambilan keputusan penanggulangan gempa bumi

Penyampaian laporan di client akan diberikan baik dengan teks atau grafik dan dapat dilihat dengan bermacam variasi dan multi dimensi. Teknik Drill-down, Roll up, slice dan dice akan dipergunakan untuk menampilkan multi dimensi pada hypercubes. ETL (Extraction, Transformation and Loading) aplikasi dapat diletakkan pada server database yang otomatis akan mengekstrak data dari transaksi database dan sumber eksternal kedalam Data Warehouse. Untuk peningkatan performance dan tidak mengganggu proses maka data akan ditangkap dengan metode *Deferred data extraction base on Date and time stamp* [Ponniah 2001]. Untuk Data Warehouse dimensi model akan digunakan snowflake skema. Jenis user seperti *miners, explorers, farmers, tourists and operators* dapat menggunakan fasilitas Data Warehouse ini [Ponniah 2001].

Data Warehouse untuk sistem pengambilan keputusan untuk penanggulangan bencana alam gempa bumi ini akan merekam lokasi bencana, jumlah korban, infrastruktur dan property yang hancur [Presiden 2008b]. Tabel Quake adalah fact tabel yang akan merekam tanggal, waktu, longitudinal, latitude, magnitude, epicenter, length, dan are gempa. Tabel lainnya adalah dimensional tabel. Tabel Quake akan mempunyai koneksi dengan tabel Stasiun Quake sebagai tabel yang mendaftar earthquake station[Geophysics].

Untuk mempermudah perekaman data area gempa bumi maka tabel quake akan dihubungkan dengan tabel regency dan provinces tabel sebagai data kabupaten dan propinsi di Indonesia. Untuk merekam jumlah korban maka tabel dead akan dihubungkan dan juga jumlah korban luka akan direkam pada injured tabel. Tabel dead dan injured akan dihubungkan ke tabel medic sebagai daftar team kesehatan yang bertanggung jawab pada korban meninggal dan luka. Tabel dead, injured dan medic akan dihubungkan ke tabel people sebagai database warga Negara.

Tabel people adalah sebagai sebuah improvement untuk kemudahan perekaman dan pengelolaan data warga Negara. Saatnya Indonesia harus mempunyai sebuah database terpusat dan setiap data mewakili seorang warga Negara dan tidak boleh seorang warga memiliki lebih dari satu identitas. Satu identitas untuk satu warga akan mempermudah pemerintah mengontrol warganya, mudah mengontrol orang-orang jahat, teroris yang merusak nama baik bangsa dan tidak akan pernah lagi seorang warga mempunyai lebih dari satu identitas. Pemerintah lebih mudah mengenali warganya, dimana mereka saat ini, dulu dan dimana mereka lahir, kapan dan diman mereka meninggal, tingkat pendidikan, siapa yang berhak untuk mendapatkan tunjangan. Siapa yang layak untuk menjadi anggota wakil rakyat, siapa yang curang dan mempunyai dua suara dalam pemilihan umum. Banyak hal yang dapat dilakukan akan tetapi pelaksanaannya adalah tergantung pemerintah sendiri.

Gambar 4. Desain entity class diagram Data Warehouse

Tabel building adalah tabel yang menggambarkan kerusakan infrastruktur, rumah, kantor, sekolah, rumah sakit dan puskesmas, tempat-tempat umum dan semua tipe peruntukan bangunan. Tabel Quake, Dead, Injured dan Building merupakan Data Warehouse utama untuk sistem pengambilan keputusan untuk penanggulangan bencana alam gempa bumi dan tabel lainnya sebagai eksternal tabel yang berasal dari institusi pemerintah seperti pemerintah daerah dan departemen. Tabel Stasiun Quake didapat dari sistem informasi BMG, tabel regency dan province dari sistem informasi administrasi nasional. Tabel people sebagai data warga negara berasal dari sistem informasi national demographi dan tabel medic berasal dari sistem informasi departemen kesehatan. Kesimpulannya sistem pengambilan keputusan untuk penanggulangan bencana alam gempa bumi ini akan dihubungkan dengan sistem informasi sub pemerintah lainnya seperti departemen dan pemerintahan daerah.

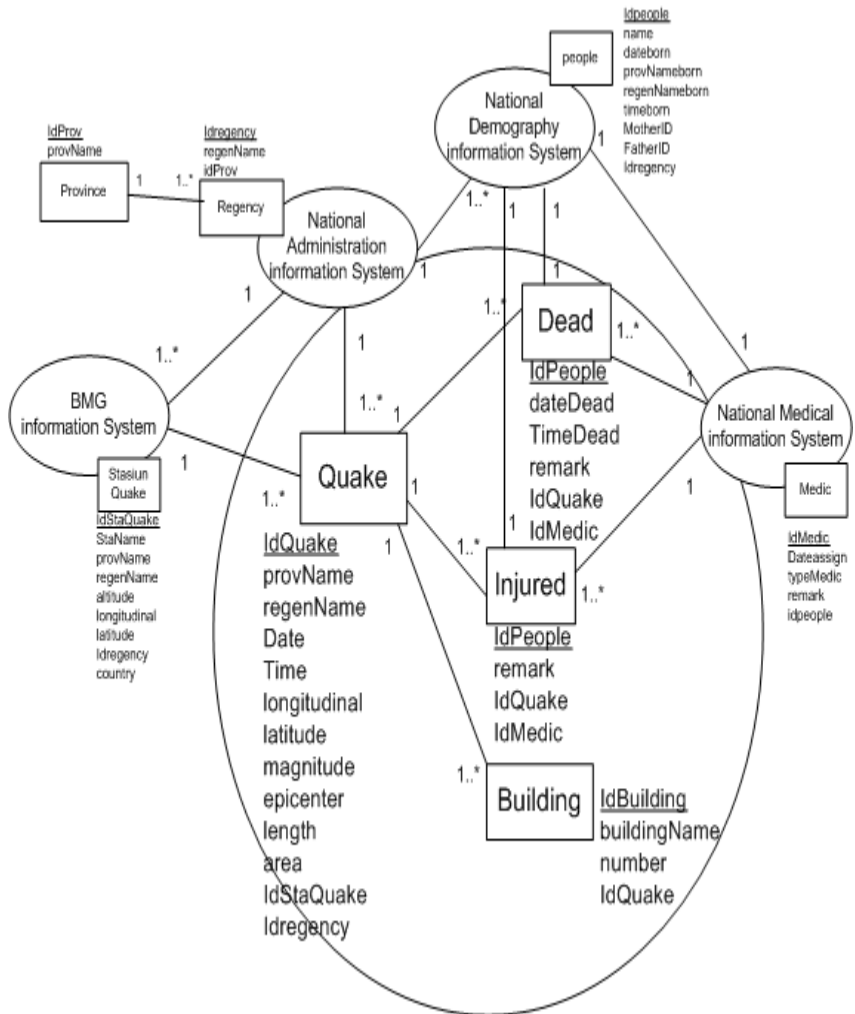

Gambar 5. Data Warehouse dan konektivitasnya dengan institusi sub pemerintah lainnya

Data Warehouse sistem pengambilan keputusan untuk penanggulangan bencana alam gempa bumi akan menjadi *Data Mart* jika ada Data Warehouse lainnya di institusi sistem informasi sub pemerintah lainnya dan tentu saja untuk kebaikan bersama akan lebih baik jika setiap sub institusi pemerintahan mempunyai masing-masing Data Warehouse dan menjalankan fungsi *EGorvernment* sebagai penghargaan dan pelayanan kepada masyarakat. Sistem pengambilan keputusan untuk penanggulangan bencana alam gempa bumi akan lebih solid juga jika ada sistem pengambilan keputusan untuk penanggulangan bencana alam lainnya yang tergabung dalam sistem manajemen penanggulangan bencana nasional. Sebagai contoh sistem pengambilan keputusan untuk penanggulangan bencana alam lainnya seperti :

1) Sistem pengambilan keputusan untuk penanggulangan bencana alam Tsunami.
2) Sistem pengambilan keputusan untuk penanggulangan bencana kegagalan Teknologi.
3) Sistem pengambilan keputusan untuk penanggulangan bencana alam kebakaran hutan.
4) Sistem pengambilan keputusan untuk penanggulangan bencana alam angin puyuh dan badai.
5) Sistem pengambilan keputusan untuk penanggulangan bencana alam letusan gunung berapi.
6) Sistem pengambilan keputusan untuk penanggulangan bencana alam kekeringan.
7) Sistem pengambilan keputusan untuk penanggulangan bencana alam longsoran.
8) Sistem pengambilan keputusan untuk penanggulangan bencana alam gelombang pasang.
9) Sistem pengambilan keputusan untuk penanggulangan bencana alam banjir.

10) Sistem pengambilan keputusan untuk penanggulangan bencana penyakit menular.
11) Dan lain-lain.

Manajemen penanggulangan bencana alam nasional akan mempunyai hubungan dengan sub institusi pemerintah lainnya, dengan dunia internatsional dan bahkan dengan pemerintah luar.

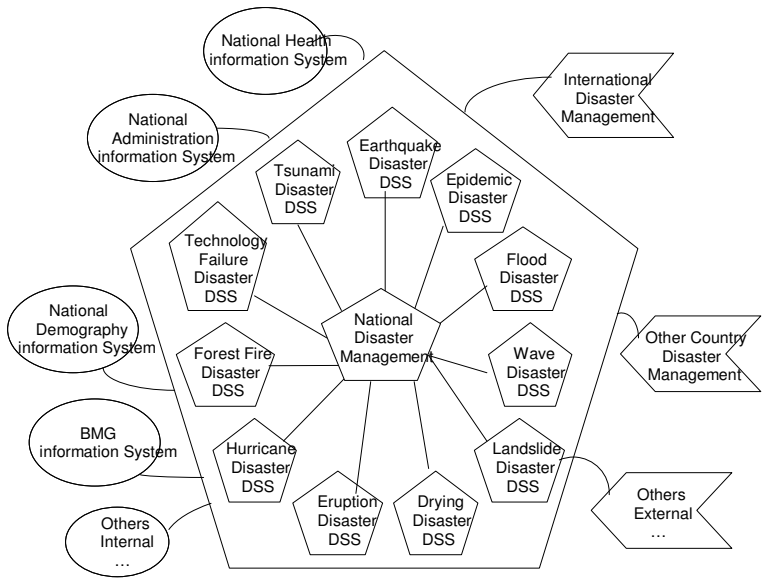

Gambar 6. Manajemen penanggulangan bencana dan hubungannya dengan internal dan eksternal aktor.

## 4. SKENARIO SISTEM PENGAMBILAN KEPUTUSAN UNTUK PENANGGULANGAN BENCANA ALAM GEMPA BUMI

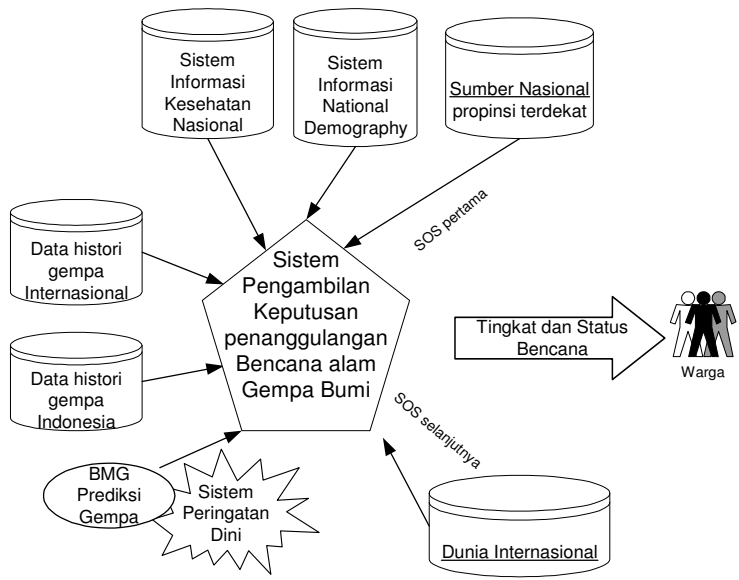

Gambar 7. Skenario Sistem Pengambilan Keputusan untuk penanggulangan bencana alam Gempa Bumi

Pertanyaan lanjutan bagaimana kira-kira Sistem Pengambilan Keputusan untuk penanggulangan bencana alam Gempa Bumi ini akan dijalankan ? Skenarion dibawah ini akan menjelaskan apa yang tergambar pada gambar 7 diatas bagaimana Sistem Pengambilan Keputusan untuk penanggulangan bencana alam Gempa Bumi akan dijalankan.

1. Sistem Pengambilan Keputusan untuk penanggulangan bencana alam Gempa Bumi akan mendapatkan peringatan sesuai dengan keluaran dari sistem peringatan dini dari BMG prediksi gempa. Peringatan akan diberikan oleh BMG berdasarkan gempa yang beresiko besar, yang mempunyai skala righter yang tinggi atau gempa yang akan menghantam lokasi-lokasi strategis dan berpenduduk banyak.

2. Berdasarkan prediksi gempa BMG, Sistem Pengambilan Keputusan untuk penanggulangan bencana alam Gempa Bumi akan mendapatkan data-data dari histori gempa bumi di Indonesia dan internasional dan membuat perbandingan. Berdasarkan informasi tersebut dan perbandingannya maka Sistem Pengambilan Keputusan untuk penanggulangan bencana alam Gempa Bumi akan memberikan prediksi status dan tingkatan bencana. Sistem Pengambilan Keputusan untuk penanggulangan bencana alam Gempa Bumi akan memprediksi beberapa hal pertanyaan berikut ini :
   a. Berapa banyak tenaga kesehatan nasional yang dibutuhkan ?
   b. Berapa banyak tenaga kesehatan internasional yang dibutuhkan ?
   c. Berapa banyak kira-kira korban meninggal dunia berdasarkan data-data lampau ?
   d. Berapa banyak kira-kira korban luka-luka berdasarkan data-data lampau ?
   e. Berapa banyak tenaga sukarela nasional yang dibutuhkan ?
   f. Berapa banyak tenaga sukarela internasional yang dibutuhkan ?
   g. Berapa tenda-tenda yang dibutuhkan ?
   h. Berapa banyak lokasi pengungsian yang dibutuhkan untuk menghindarkan kepadatan pengungsi dalam sebuah lokasi ?
   i. Lokasi mana yang strategis sebagai tempat pengungsian ?
   j. Berapa banyak tempat sanitasi yang harus dibangun ?
   k. Berapa banyak tempat dapur umum yang harus dibangun ?
   l. Berapa banyak kilogram beras yang harus disediakan ?
   m. Sumber-sumber makanan mana dan yang terdekat yang dapat diakses ?
   n. Berapa banyak makanan bayi yang harus disediakan ?
   o. Berapa banyak selimut yang harus disediakan ?
   p. Berapa total biaya yang dibutuhkan ?
   q. Perkiraan total kerugian dan kerusakan dalam satuan angka bangunan ?
   r. Perkiraan total kerugian dan kerusakan dalam satuan mata uang ?
   s. Dan pertanyaan-pertanyaan lainnya.
   Dan diantaranya hal-hal yang harus mendapatkan perhatian lebih adalah [Presiden 2008b] adalah makanan, pakaian, air, sanitasi, team penyelamat, pelayanan kesehatan dan pelayanan psikologi.

3. Berdasarkan prediksi gempa BMG, Sistem Pengambilan Keputusan untuk penanggulangan bencana alam Gempa Bumi akan mendapatkan informasi mengenai prediksi jumlah masyarakat pada area gempa dari sistem informasi demografi nasional. Untuk kepentingan rumus maka satuan W= Jumlah warga pada area gempa.

4. Berdasarkan prediksi gempa BMG, Sistem Pengambilan Keputusan untuk penanggulangan bencana alam Gempa Bumi akan mendapatkan informasi mengenai prediksi jumlah tenaga kesehatan pada area gempa dari sistem informasi kesehatan nasional. Berdasarkan data standar nasional penanganan maksimal warga untuk setiap tenaga kesehatan pada waktu bencana yang dikeluarkan oleh sistem informasi kesehatan nasional. Tenaga kesehatan yang dibutuhkan akan didapatkan dari jumlah warga pada area gempa (W) dan dibagi dengan standar nasional maksimal penanganan warga per setiap tenaga kesehatan pada waktu bencana. Untuk kepentingan rumus maka Sn = standar nasional maksimal penanganan warga per setiap tenaga kesehatan pada waktu bencana, Tk= Tenaga kesehatan yang dibutuhkan dan Jtk=Jumlah tenaga kesehatan pada area gempa

Rumus : $$Tk = W / Sn \qquad\qquad (1)$$

5. Jika tenaga kesehatan yang dibutuhkan (Tk) lebih dari jumlah tenaga kesehatan pada area gempa (Jtk) maka kekurangan tenaga kesehatan akan didapatkan dengan mengurangkan tenaga kesehatan yang dibutuhkan(Tk) dengan jumlah tenaga kesehatan pada area gempa(Jtk). Untuk kepentingan rumus maka Ktk=Kekurangan tenaga kesehatan.

Rumus: $$\text{if } (Tk{>}Jtk) \ Ktk = Tk \text{ -}Jtk \qquad\qquad (2)$$

6. Jika Kekurangan tenaga kesehatan (Ktk) lebih dari 0 maka Sistem Pengambilan Keputusan untuk penanggulangan bencana alam Gempa Bumi akan menyatakan SOS pertama sebagai tanda permintaan untuk permintaan bantuan dari sumber dan cadangan nasional, propinsi atau pemerintahan daerah yang terdekat [Presiden 2008b] yang dapat membantu menyediakan tenaga kesehatan. Jumlah tenaga kesehatan yang dapat disediakan pada waktu bencana dari setiap propinsi dan kabupaten akan distandarisasikan di sistem informasi kesehatan nasional yang dikelola oleh departemen kesehatan.

7. Setelah mendapatkan jumlah tenaga kesehatan yang dapat membantu dari sumber dan cadangan nasional seperti propinsi dan pemerintahan daerah yang terdekat dan jika masih saja minus maka Sistem Pengambilan Keputusan untuk penanggulangan bencana alam Gempa Bumi akan mengirimkan SOS kedua sebagai tanda permintaan bantuan kepada dunia internasional [Presiden 2008c]. Seperti pada bencana alam gempa bumi yang sudah-sudah, korban bertambah karena pemerintah telat untuk menyatakan permintaan bantuan kepada dunia internasional sebagai bentuk gengsi pemerintah untuk menadahkan tangan, padahal rakyat sangat membutuhkan dan lagipula banyak dunia internasional sudah terlebih awal menawarkan bantuan terlepas dari kepentingan masing-masing organisasi internasional tersebut.

Dengan membuat Sistem Pengambilan Keputusan untuk penanggulangan bencana alam Gempa Bumi lebih jelas dan terorganisasi maka akan membantu pemerintah untuk dapat cepat dan cakap dalam membuat keputusan penanggulangan bencana alam gempa bumi didukung oleh informasi dan data yang akurat dan dapat dipercaya. Semakin lengkap data-data yang disediakan maka keakuratan informasi sebagai masukan untuk membuat keputusan akan lebih terjamin.

## 5.   KESIMPULAN

Tulisan ini merupakan sebagian dari apa yang diamanatkan oleh undang-undang dan peraturan pemerintah bahwa pada waktu tidak ada bencana maka riset dan penelitian yang berhubungan dengan manajemen bencana dapat dilakukan untuk mendukung manajemen bencana [Presiden 2008b].

Sulit dilakukan implementasi jika tidak adanya dukungan dan keinginan pemerintah untuk melaksanakannya dan pemerintah yang bersih, transparansi dan bertanggung jawab akan sangat besar pengaruhnya.

Penelitian untuk bencana-bencana lainnya akan sangat dibutuhkan sebagai bagian dari keseluruhan sistem manajemen bencana nasional.

Simulasi dengan pendekatan komputer game akan lebih meningkatkan minat dan nilai tambah untuk Sistem Pengambilan Keputusan untuk penanggulangan bencana alam Gempa Bumi didalam mendidik pembuat keputusan, para peneliti dan warganya. Dengan adanya unsur komputer game sebagai hal yang menarik akan dapat menarik juga penelitian tambahan dan minat dari kaum muda yang senang dengan game sebagai hal yang menyenangkan, menghibur dan tidak membosankan.

Implementasi sistem informasi geographi pada Sistem Pengambilan Keputusan untuk penanggulangan bencana alam Gempa Bumi akan menyedikan alat yang ampuh untuk pemecahan permasalahan tertentu dan masalah-masalah yang menarik di bidang seismologi [Phuong 2004].

Penelitian tentang pengenalan wajah ('*face recognition*') dapat dikembangkan untuk sebuah sebuah satu identitas untuk setiap warga negara, dimana setiap orang dapat dikenal dari wajahnya, dimana wajah manusia akan disimpan didalam database dan dapat dikenal pada waktu dilakukan pencarian berdasarkan data-data gambar/video yang berisi wajah manusia. Pengenalan dengan wajah manusia akan lebih menjamin satu identitas untuk setiap warga negara.

# REFERENSI